\documentclass[12pt]{article}
\usepackage{amsfonts}
\usepackage{amsmath}
\usepackage[T2A]{fontenc}
\usepackage[cp866]{inputenc}
\usepackage[dvips]{graphicx}
\usepackage{epsfig}
\usepackage{amssymb}

\begin{document}
\def\be{\begin{equation}}\def\ee{\end{equation}}\def\l{\label}
\def\0{\setcounter{equation}{0}}\def\b{\beta}\def\S{\Sigma}\def\C{\cite}
\def\r{\ref}\def\ba{\begin{eqnarray}}\def\ea{\end{eqnarray}}
\def\n{\nonumber}\def\R{\rho}\def\X{\Xi}\def\x{\xi}\def\la{\lambda}
\def\d{\delta}\def\s{\sigma}\def\f{\frac}\def\D{\Delta}\def\pa{\partial}
\def\Th{\Theta}\def\O{\Omega}\def\o{\omega}\def\th{\theta}\def\ga{\gamma}
\def\Ga{\Gamma}\def\t{\times}\def\h{\hat}\def\rar{\rightarrow}
\def\vp{\varphi}\def\inf{\infty}\def\le{\left}\def\ri{\right}
\def\foot{\footnote}\def\ve{\varepsilon}\def\N{\bar{n}(s)}\def\cS{{\cal S}}
\def\k{\kappa}\def\sq{\sqrt{s}}\def\bx{{\bf x}}\def\La{\Lambda}
\def\bb{{\bf b}}\def\bq{{\bf q}}\def\cp{{\cal P}}\def\tg{\tilde{g}}
\def\cf{{\cal F}}\def\bN{{\bf N}}\def\Re{{\rm Re}}\def\Im{{\rm Im}}
\def\K{{\cal K}}\def\ep{\epsilon}\def\cd{{\cal d}}\def\co{\hat{\cal
O}} \def\j{{\h j}}\def\e{{\h e}}\def\F{{\bar{F}}}\def\cn{{\cal N}}
\def\P{\Phi}\def\p{\phi}\def\cd{\cdot}\def\L{{\cal L}}\def\U{{\cal U}}
\def\Z{{\cal Z}}\def\ep{\epsilon}\def\a{\alpha}\def\ru{{\rm u}}
\def\vep{\varepsilon}\def\ix{\mathbf{x}}\def\vs{\varsigma}\def\z{\zeta}
\def\vr{\varrho}\def\Sp{\mathrm{Sp}}
\begin{center}
{\Large\bf Phenomenology of Very High Multiplicity Hadron Processes}
\vskip 0.5cm {\large\it J.Manjavidze\foot{JINR,Dubna, Russia $\&$
Inst. of Phys., Tbilisi, Georgia. E-mail: joseph@nusun.jinr.ru} and
A.Sissakian\foot{JINR, Dubna, Russia. E-mail: sisakian@jinr.ru}}
\end{center}

\begin{abstract}

We discuss the possibility to suppress the nonperturbative effects if
the very high multiplicity hadron final states are chosen. The
theoretical uncertainties and possible experimental measurements are
described.
\end{abstract}

\vskip 0.5cm {\bf\large 1.} Very high multiplicity (VHM) events are
expected to be produced at the future experiments at BNL (STAR),
Fermilab (CDF), CERN (ATLAS). Then, it is natural to expect the VHM
region to give new important information, which will be complementary
to the ongoing investigations and/or unavailable in other than the
VHM process. The studies of the VHM region have been proposed
recently in \C{1} and attract an increasing interest in high energy
physics \C{vhm, drem, ichep, jenk}.

In this Letter, where we briefly consider the VHM topic in whole, we
particularly emphasize the related physical quantities expected to be
measured with forthcoming experiments. In this sense, it is crucial
to understand experimental abilities of such measurements, since the
topological cross section, $\s_n$, is extremely small (e.g., at TeV
energies, $\s_n\ll 10^{-7}\s_{tot}$, compared to the total cross
section $\s_{tot}$, Fig.1) and could lead to certain
problems.\medskip

\begin{figure}[thb]
\begin{center}
\begin{tabular}{c}
\mbox{\epsfig{figure=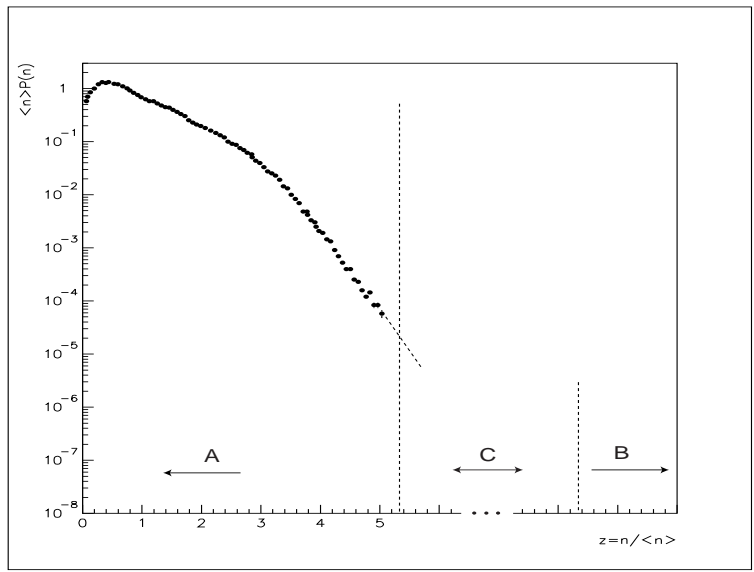,
width=0.6\textwidth,height=0.3\textheight}}
\end{tabular}
\end{center}\vskip -0.5cm
\caption{\footnotesize The KNO multiplicity distribution,
$P(n)=\s_n/\s_{tot}$. The points are the E735 Tevatron data \C{teva}.
The range {\bf A} corresponds to the multiperipheral kinematics. The
range {\bf B} corresponds to the non-interacting, ideal, gas
approximation, $n\ll n_{max}.$ {\bf C} is the VHM region.}
\end{figure}

Started investigations of VHM processes, one has first of all ask a
question: why the mean multiplicity in hadron processes being energy
dependent as $\ln^2(s/m_\pi^2)$ is much $smaller$ than the threshold
multiplicity value $n_{max}=\sqrt s/m_\pi$ \C{mea}? Here $\sqrt s$ is
the c.m.s. energy and $m_\pi$ is the pion mass.

In the present Letter, we follow the idea \C{1} that the multiple
production process could be considered as a process of hot tea
cooling in a cold environment. Then very hot tea cooling in very cold
room is analogous to the VHM process. In that case, the cooling
process proceeds "quickly", i.e. any process obstacled to the cooling
fluctuations must be suppressed. The idea that the hadron multiple
production process may consist of various mechanisms has been
considered in \C{siss}.

In the QCD parton picture \C{qcdp} the "fastness" of a process means
that parton decay must prevail over its dissipation from an
interaction zone. This is possible if the parton virtuality $|q|$ is
high enough as far as the parton lifetime is $\sim1/|q|$. As a
result, the VHM process should be hard and, therefore, the forces
preventing particle production (confinement) must be negligible.

Fig.2 shows the longitudinal-transverse momenta phase space domain of
the VHM process. It is remarkable that the VHM domain occupies both
regions, the "low-$p_\parallel$" and the "low-$p_\bot$" ones.
Therefore, despite the fact that the VHM process is hard one can not
use the Leading Logarithm Approximation (LLA) \C{lla} to describe it
\C{1}. To stress is that, following the LLA formalism, there are the
only two domains, namely the Regge and Deep inelastic Scattering
(DIS) domains, as shown in Fig.2. The corresponding cross sections
may be sufficiently large and the LLA is applicable. \medskip

\begin{figure*}[tbph]
\begin{center}
\begin{tabular}{c}
\mbox{\epsfig{figure=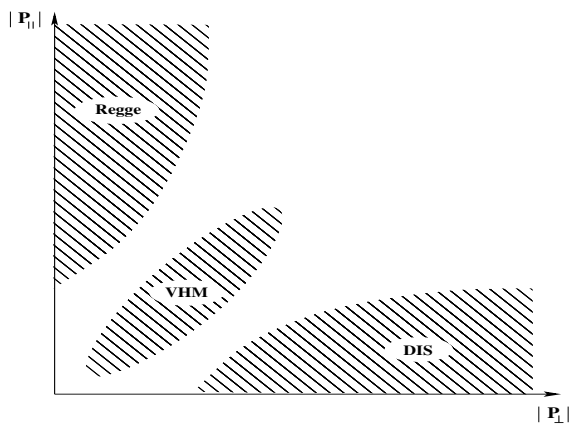,
width=0.45\textwidth,height=0.22\textheight}}
\end{tabular}
\end{center}
\vskip -0.8cm \caption{\footnotesize The longitudinal-transverse
momentum phase space of inelastic hadron processes, see text. }
\end{figure*}

The "low-$p_\parallel$", or  "low-$x$" \C{lev} in DIS, also means
that VHM particles have low energies. For that reason, the hadrons
are mainly resulted from the "gluing" of coloured constituents
created in the hard process. Therefore, {\it the (soft) process of
coloured parton pair production from the vacuum is negligible in the
VHM domain}. This distinguishes VHM processes from other hard
processes (e.g., from hadron production in $e^+e^-$-annihilations, or
DIS processes). The VHM dynamics is free of the non-perturbative
background.

Sec.2 argues the above conclusion, in Sec.3 the main properties of
VHM dynamics are discussed and the concluding remarks are given in
Sec.4.\medskip

{\bf\large 2.} The dominance of "Regge" processes stands for the
"softness" of hadron dynamics \C{landsh}. This is based on the
experimental observations: the mean transverse momentum of secondary
particles, being energy and multiplicity independent, has a limited
value, and the diffraction radius of hadron elastic scattering
increases with increasing energy. Such "multiperipheral" kinematics
is described by the Regge model with the {\it soft Pomeron} ansatz.

From the "microscopical" point of view, it means that the soft
process can be roughly described by the very spatial class of the
Feynman "ladder" diagrams of perturbative QCD (pQCD). This is the
so-called "{\it BFKL Pomeron}" solution \C{kur}. It is based on the
LLA approach and assumes that the process described is sufficiently
hard\foot{The point is that this Regge-like theory depends on the
phenomenological energy scale parameter $s_0$ and, therefore, its
range of application may be chosen arbitrary.}.

The both models are based on the assumption that the mean transverse
momentum of produced hadrons is small. In the frame of the {\it soft
Pomeron} approach, this constrain is hidden in the conservation laws,
which are the consequence of the underlying non-Abelian gauge
symmetry. In the {\it BFKL Pomeron} case, the LLA place a role of the
constrain.

It is worse to note that the constrain restricts the production
dynamics and, therefore, the produced states can not entirely cover
the phase space with the same density, Fig.2. This requires the
constrain to be suppressed in order to lead to the VHM final states.
Remembering that the constrains are the consequence of the long-range
connections among interacting degrees of freedom, this also points
out that the VHM process must be the hard process.

The VHM scenario discussed realizes if (i) the hard processes has
place in the VHM region and if (ii) the hard process ensures the
"fast" hadron production.

The formal proof of the item (i) contains the following steps
\C{1,4}. First, it can be shown that if the final-state interactions
are excluded, then the $soft$ process only satisfies the following
asymptotics:
$$\s^s_n< O(e^{-n}),$$ i.e. the "soft" topological cross section
$\s^s_n$ falls down faster than any power of $\exp\{-n\}$. The reason
is just the softness of the process. On the other hand, the $hard$
process leads to the following asymptotic estimate:
$$\s^h_n= O(e^{-n}).$$ Consequently, one always finds such energy
$\sqrt s$ and multiplicity $n$ that
$$\s^h_n\gg\s^s_n~{\rm for}~n\ll\sqrt{s} /m_\pi.$$ This proves
the item (i). \medskip

\begin{figure*}[tbph]
\begin{center}
\begin{tabular}{c}
\mbox{\epsfig{figure=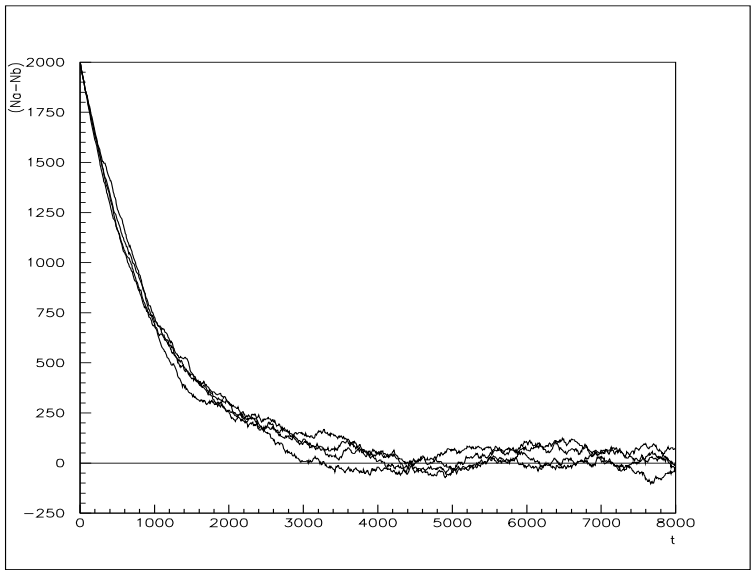,
width=0.6\textwidth,height=0.3\textheight}}
\end{tabular}
\end{center}\caption{\footnotesize The Ehrenfest-Kac model simulation for random ball
shifting from the $a$ box to the $b$ box. Initially the number of
balls is $N_a$=2000 and $N_b$=0, while in the equilibrium,
$N_a-N_b=0$. Four simulations are displayed.}
\end{figure*}

The qualitative argument for the statement (ii) follows from the
Ehrenfest-Kac model \C{5} of the irreversibility phenomenon\foot{The
Ehrenfest-Kac model considers two box $a$ and $b$ and $2N$ $numbered$
balls. One may assume that the "far from the equilibrium" initial
state presents the situation when all $2N$ balls are e.g., in the box
$a$ and in the equilibrium each box contains $N$ balls. Then,
choosing $randomly$ the number of balls, one should take the
corresponding ball from the box and put it into the other one. At the
following step, the number of a ball will be again randomly chosen,
and the corresponding ball should be put into the other box, and so
on. One may measure the time $t$ in the units of the ball shifting
time. Then, the number of the "produced particles" in the box $b$
measures the trend to equilibrium. In this equilibrium state,
$N_a=N_b$ (Fig3).}. It shows that if the initial state is {\rm far}
from the final one, then the system undergoes to equilibrium {\it in
the fastest way }\foot{This phenomenon was not mentioned in the known
publications concerning the Ehrenfest-Kac model.}. The result of the
computer modelling of "particle production" within discussed model,
Fig.3, confirms this. Indeed, as it is seen there are no fluctuations
at early stage of the process: number of the "produced particles" is
proportional to $t$, up to $t\simeq1000$.

The formal proof of the statement (ii) can be obtained from the
KNO-scaling of the multiplicity distribution in pQCD jets \C{1}.
Introducing the quantity \be
\mu(n,s)=-\f{1}{n}\ln\le\{\f{\s_n(s)}{\s_{tot}(s)}, \ri\}\l{1}\ee one
finds at the mass $M$ of a jet, $$ \mu_1(n,M)= \f{a}{\bar{n}_j(M^2)}+
O(1/n).$$ Here $a>0$ and $\bar{n}_j(M^2)$ is the mean hadron
multiplicity in a QCD jet, $\ln\bar{n}_j(M^2)\sim\{\ln
(M^2/\La^2)\}^{1/2}$. The two-jet contribution at the same energy $M$
reads as
$$\mu_2 (n,M)=\f{a}{\bar{n}_j(s/4)}+ O(1/n)>\mu_1(n,s),$$ where the
energy conservation law is taken into account. Therefore, the
asymptotics over $n$ is defined by the events with the lowest numbers
of pQCD jets.

This confirms the statement (ii) since just the jet mechanism of
particle production is known to be the fastest one. Moreover, we
conclude that the number of the jets decreases with increasing total
multiplicity $n$ and, thus, the heavier jets are produced in the VHM
region. However, one has to note that this conclusion is based on the
LLA and, therefore, we consider it just as a qualitative
prediction.
\medskip

{\bf\large 3.} From the above, one concludes that the VHM process is
a hard process. In this case, the process of incident energy
dissipation evolves $freely$, i.e. without the influence of the
"hidden constrains". This is the important observation because it
allows to assume that in such conditions the of final state
distribution trends to the "equilibrium".

Let us remind some statistical properties of the maximally
constrained systems, also called the "exactly integrable systems". It
is known that if the system has $2N$ degrees of freedom and holds $N$
constrains (first integrals in involution) then no thermalization
occurs in such a system. The first observation of this phenomenon
belongs to Fermi, Pasta and Ulam \C{1,7}. On the other hand, it is
evident that the system without any constrains should drift to the
equilibrium.

The hadron dynamics hides definite number of constrains, which is not
enough to completely suppress particle production, i.e. the
thermalization process. This conclusion follows from the existence of
large multiplicity fluctuations, see Fig.1. Nevertheless, the
constrains occur and the mean multiplicity is reasonably small $\N\ll
n_{max}$.

To stress is that the term "equilibrium" is understood here as the
absence of any macroscopic correlations among the extensive
parameters of a system. For example, the absence of energy
correlations implies the thermal equilibrium. Then, one can obtain
the experimental condition, which reflects the "equilibrium
phenomena" \C{1,8}. It has been shown that the condition \be
R_l(n,s)=\f{|K_l(n,s)|^{2/l}}{|K_2(n,s)|}<1\l{2},~~l=3,4,...,\ee is
the necessary and sufficient one to reach thermalization. Here
$K_l(n,s)$ are the $l$-point energy correlators in the $n$-particle
system, $l<n$. To note is that (\r{2}) reminds the "correlation
relaxation condition" of Bogolyubov \C{bogo}.

The tendency of the system to reach the equilibrium is considered as
the dynamical effect, when the (symmetry) constrains  are switched
off, and for that reason the system may drift to the equilibrium
state.

\begin{figure*}[tbph]
\begin{center}
\vskip -0.5 cm
\begin{tabular}{c}
\mbox{\epsfig{figure=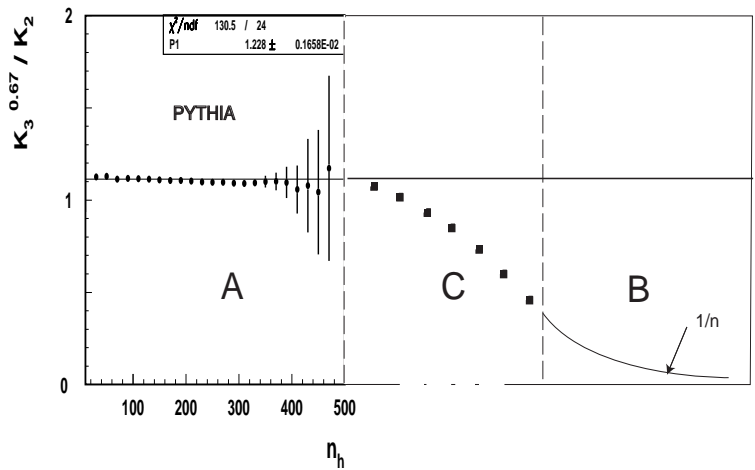,
width=0.6\textwidth,height=0.3\textheight}}
\end{tabular}
\end{center}
\vskip -1cm \caption{\footnotesize The ratio
$R_3(n,s)=\le|K_3\ri|^{2/3}(n,s)/\le|K_2(n,s)\ri|$ of the energy
correlators vs. multiplicity. The PYTHIA prediction is given for the
{\bf A} domain (see Fig.1). Domain {\bf B} corresponds to the
"non-interacting gas" approximation.}
\end{figure*}

Fig.4 shows the result of Monte Carlo PYTHIA event generator \C{pyth}
simulation of the ratio $R(n,s)$ for the domain {\bf A} in Fig.1. The
absence of the thermalization process is natural if one notes that
PYTHIA resembles the Regge model. At the same time, it can be shown
that in the region {\bf B} the system without fail reaches the
thermal equilibrium, $R(n)\sim 1/n$ in that domain.
\medskip

{\bf\large 4.} We can conclude that:

(i) the VHM process allows us to investigate the hadron dynamics
beyond the standard multiperipheral kinematics;

(ii) it is the hard process in sense that the influence of hidden
constrains, connected with the high symmetries and the nontrivial
vacuum of the Yang-Mills theory, is not important in the particle
production dynamics;

(iii) the VHM process allows us to reach the thermal equilibrium,
where the ordinary thermodynamical methods can be used. (This extends
experimental possibilities and, therefore, is extremely important.
For example, the $\mu$-quantity introduced in (\r{1}) is the
"chemical potential" if measured in the units of the mean energy of
produced particles \C{1});

(iv) the VHM domain seems to be one of the mostly interesting regions
to investigate the collective phenomena, e.g. phase transitions. (To
note is that the VHM final state is a "cold" one).

To note is that the production dynamics of VHM states might be well
enough described by pQCD, but the logarithmic accuracy seems to be
not enough. This is a main theoretical problem and to this end a more
accurate perturbation theory, described in \C{9}, has been
formulated.

At the end, let us add that, as it follows from the above conclusion,
the VHM process leads to the $dense$, $cold$ and $equilibrium$
locally-coloured state, e.g. "cold" plasma \C{coldpl}\foot{It is hard
to imagine the coloured state to form the QCD plasma one if the
former is "hot". In this case the kinetic motion would rapidly
separate the charges and this must lead to the strong polarization of
the vacuum. This effect is absent in the QED plasma.}.
\medskip

\noindent{\Large \bf Acknowledgements}\footnotesize{

Authors would like to take the opportunity to thank Yu. Budagov,
V.Kadyshevski, A.Korytov, E.Kuraev, L.Lipatov, V.Matveev, V.Nikitin
and E.Sarkisyan for fruitful discussions and constructive comments.
Special thanks to STAR (BNL) and CDF (FNAL) physics communities for
continuous interest in the VHM problem.

}

\end{document}